\def\rn{\noindent\parshape 2 0truecm 8.5truecm 0.3truecm 8.2truecm}
\def\rn{}
\def\nn#1 #2{#2. #1}				
\def\nnn#1 #2 #3{#2. #3. #1}			
\def\nnnn#1 #2 #3 #4{#2. #3. #4 #1}		
\def\nnnnn#1 #2 #3 #4 #5{#2. #3. #4 #5. #1}	
\def\dualand{ and\hbox{ }}				
\def\multiand{, and\hbox{ }}				
\def\rf#1;#2;#3;#4;#5 {{\frenchspacing\par\rn#1, #3 {\bf #4}, #5 (#2). \par}}
\def\rg#1;#2;#3;#4;#5;#6 {{\frenchspacing\par\rn#1, #3 {\bf #4}, #5 (#2). \par}}
\def\rfbook#1;#2;#3;#4;#5 {{\frenchspacing\par\rn#1, {\it #3} (#5, #4, #2).\par}}
\def\rfprep#1;#2;#3 {{\par\frenchspacing\rn#1, #3 (#2).\par}}

\def\expec#1{\langle#1\rangle}

\def\etal{{\frenchspacing\it et al.}}

\def\eg{{\frenchspacing\it e.g.}}
\def\etc{{\frenchspacing\it etc.}}

\def\beq#1{\begin{equation}\label{#1}}
\def\eeq{\end{equation}}
\def\beqa#1{\begin{eqnarray}\label{#1}}
\def\eeqa{\end{eqnarray}}
\def\eq#1{equation~(\ref{#1})}

\def\fig#1{Figure~\ref{#1}}
\def\Fig#1{Figure~\ref{#1}}

\def\sec#1{Section~\ref{#1}}

\def\spose#1{\hbox to 0pt{#1\hss}}
\def\simlt{\mathrel{\spose{\lower 3pt\hbox{$\mathchar"218$}}
     \raise 2.0pt\hbox{$\mathchar"13C$}}}
\def\simgt{\mathrel{\spose{\lower 3pt\hbox{$\mathchar"218$}}
     \raise 2.0pt\hbox{$\mathchar"13E$}}}
\def\simpropto{\mathrel{\spose{\lower 3pt\hbox{$\mathchar"218$}}
     \raise 2.0pt\hbox{$\propto$}}}

\def\ed{\end{document}}

\def\draft{
}

\def\l{\ell}

\def\p{{\bf p}}

\def\F{{\bf F}}

\def\SS{{\bf S}}

\def\W{{\bf W}}

\def\lmax{\l_{\rm max}}
\def\zz#1#2#3{$#1^{+#2}_{-#3}$}
\def\zzp#1#2{$#1^{+#2}$}

\def\Ob{\Omega_{\rm b}}
\def\Oc{\Omega_{\rm cdm}}
\def\Od{\Omega_{\rm dm}}
\def\Ok{\Omega_{\rm k}}
\def\Ol{\Omega_\Lambda}
\def\Om{\Omega_{\rm m}}
\def\On{\Omega_\nu}
\def\ob{\omega_{\rm b}}

\def\od{\omega_{\rm dm}}

\def\fn{f_\nu}

\def\ns{n_s}
\def\nt{n_t}
\def\As{A_s}

\def\zion{z_{\rm ion}}

\def\l{\ell}

\def\Pl{P_\delta^{\rm l}(k)}

\def\kl{k_{\rm l}}
\def\knl{k_{\rm nl}}
\def\neff{n_{\rm eff}}
\def\Deltal{\Delta_{\rm l}}
\def\Deltanl{\Delta_{\rm nl}}

\documentstyle[prd,aps,epsf]{revtex}
\draft
\begin{document}
\twocolumn[\hsize\textwidth\columnwidth\hsize\csname@twocolumnfalse\endcsname




\title{The last stand before MAP: cosmological parameters from lensing, CMB and galaxy clustering}

\author{Xiaomin Wang$^1$, Max Tegmark$^1$, Bhuvnesh Jain$^1$ \multiand Matias Zaldarriaga$^2$}

\address{$^1$Dept. of Physics, Univ. of Pennsylvania, Philadelphia, PA 19104;
  xiaomin@hep.upenn.edu}
\address{$^2$Dept. of Physics, New York University, New York, NY 10003}

\maketitle 


\begin{abstract} 
Cosmic shear measurements have now improved to the point where they deserve
to be treated on par with CMB and galaxy clustering data for cosmological parameter
analysis, using the full measured aperture mass variance curve rather than a
mere phenomenological parametrization thereof.
We perform a detailed 9-parameter analysis of recent lensing 
(RCS), CMB (up to Archeops) and galaxy clustering (2dF) data,
both separately and jointly.
CMB and 2dF data are consistent with a simple flat adiabatic scale-invariant 
model with $\Ol=0.72\pm 0.09$, $h^2\Oc=0.115\pm 0.013$,
$h^2\Ob=0.024\pm 0.003$, and a hint of reionization around $z\sim 8$.
Lensing helps further tighten these constraints, but reveals 
tension regarding the power spectrum normalization: including the 
RCS survey results raises $\sigma_8$ significantly and forces other 
parameters to uncomfortable values.
Indeed, $\sigma_8$ is emerging as the currently most controversial cosmological 
parameter, and we discuss 
possible resolutions of this $\sigma_8$ problem. 
We also comment on the disturbing fact that many recent analyses (including
this one) obtain error bars smaller than the Fisher matrix bound.
We produce a CMB power spectrum combining all existing experiments, and
using it for a ``MAP versus world'' comparison next month will provide a 
powerful test of how realistic the error estimates have been in the cosmology community.
\end{abstract}
\bigskip
] 


\section{Introduction}


An avalanche of precision data has revolutionized our ability to constrain 
cosmological models and their free parameters in recent 
years \cite{Lange00,boompa,Balbi00,observables,Jaffe00,Padmanabhan00,Lineweaver00,10par,concordance,Kinney01,Hannestad01,Hannestad01b,Griffiths01,Phillips01,BondConfProc,CosmicTriangle,Novosyadlyj00,Novosyadlyj00b,Durrer00,Bridle00,Turner01,Holder01,Efstathiou01,Pryke01,Stompor01,qmap3,KnoxPage00,VSA02d,CBI02a,xiaomin01,Tegmark02,Lewis02,Venkatesan02,Melch02a,Melch02b,Melch02c}.
Around 1998, cosmic microwave background (CMB) data had become good enough to 
allow a realistic high-dimensional parameter space to be explored by fitting 
theoretical models directly to the measured CMB power spectrum (\eg, \cite{10par,Lineweaver00}).
In contrast, other cosmological constraints were included merely as priors on individual 
parameters, for instance the baryon density from Big Bang nucleosynthesis, 
the matter density from galaxy cluster abundance and the Hubble parameter from direct observations.

With the advent of improved galaxy redshift surveys like 
PSCz \cite{Saunders00}, 2dF \cite{Percival01} and SDSS \cite{York00},
the galaxy power spectrum $P(k)$ was upgraded and admitted to the ``CMB club'':
it was so accurately measured that people started fitting models directly to the measured
$P(k)$ curve rather than merely to some phenomenological parametrization thereof (say in terms of
an amplitude and a ``shape parameter'') \cite{10par,concordance,Efstathiou01}.
The Lyman $\alpha$ forest (Ly$\alpha$F) power spectrum 
has now undergone the same upgrade
\cite{Hannestad01,xiaomin01}

One promising cosmological probe is still conspicuously absent from this club and has so far been 
left out in the cold: the cosmic shear power spectrum measured with weak gravitational lensing.
Like the CMB, it has the advantage of probing the dark matter distribution directly
\cite{Hoekstra02b},
without murky bias issues.
Despite the spectacular progress since this elusive signal was first detected in 
2000 \cite{Waerbeke00,Bacon00,Wittman00,Kaiser00}, increasing the sky area covered by almost
two orders of magnitude \cite{Hoekstra02b,RCS01a,RCS01b,Bacon02,Waerbeke01,Hoekstra02a,Refregier02,Waerbeke02,Jarvis02},
it has so far only been included in cosmological parameter studies as a prior on two parameters: 
the power spectrum normalization and the matter density.
Yet it has been shown that the detailed shape of the weak lensing power spectrum $P_\kappa(\l)$
potentially contains almost as much cosmological information as the CMB 
\cite{HuOkamoto02,Hu98,Hu00}, with great prospects for degeneracy breaking
and cross-checking.


The goal of this paper is to treat weak lensing on an equal footing with CMB and galaxy clustering,
fitting theoretical models directly to the CMB, LSS and lensing power spectra.
Our results can also be viewed as the last stand before MAP:
using our constraints combining the state-of-the-art available data for a 
``MAP versus world'' comparison next month will provide a powerful test of how realistic 
the error estimates and underlying assumptions have been in the cosmology community.

The rest of this paper is organized as follows. We present our analysis method in \sec{MethodsSec}, 
describe the CMB, galaxy and lensing data used in \sec{DataSec}, 
compute quantitative model constraints in \sec{ParameterSec} and discuss our conclusions in \sec{DiscussionSec}.

\section{Method}
In this section, we briefly describe our method for constraining cosmological models.

\label{MethodsSec}

\subsection{General method}

We use the grid-based multiparameter analysis method described in detail in 
\cite{10par,concordance,xiaomin01}, 
with modifications as described below.
It involves the following steps:
\begin{enumerate}
\item Compute CMB and galaxy power spectra and the weak lensing aperture mass variance for a grid
of models in our 9-dimensional parameter space.
\item Compute a likelihood for each model that quantifies how well it fits the
data.
\item Perform 9-dimensional interpolation and marginalization to obtain
constraints on individual parameters and parameter pairs.
\end{enumerate}
As in \cite{xiaomin01}, the we parametrize cosmology with the 11 parameters
\beq{pEq}
\p\equiv(\tau,\Ok,\Ol,\od,\ob,\fn,\ns,\nt,\As,r,b).
\eeq
They are the reionization optical depth $\tau$, 
the primordial amplitudes $\As$, $r\As$ and tilts $\ns$, $\nt$ 
of scalar and tensor fluctuations, 
the bias parameter $b$ defined as the ratio between rms 
galaxy fluctuations and rms matter fluctuations on 
large scales,
and five parameters specifying the cosmic matter budget, 
curvature $\Ok$, vacuum energy $\Ol$, cold dark matter $\Oc$, 
hot dark matter (neutrinos) $\On$ and baryons $\Ob$.
The quantities
$\ob\equiv h^2\Ob$ and
$\od\equiv h^2\Od$ correspond to 
the physical densities of baryons
and total (cold + hot) dark matter 
($\Od\equiv\Oc+\On$), and $\fn\equiv\On/\Od$ is the fraction
of the dark matter that is hot.
We assume that the galaxy bias $b$ is constant on large scales 
\cite{Taruya01}.
We make $\As$ as a discretized parameter to marginalize and find a best fit of
$\sigma_8$ later in computation, as comparing to an undiscretized $\As$ in previous papers\cite{xiaomin01}.
Throughout this paper, we assume a negligible contribution from spatial curvature and massive 
neutrinos, setting $\Ok=\fn=0$.
Our final parameter grids are as follows:
\begin{itemize}
\item $\tau=0, 0.05, 0.1, 0.2, 0.3, 0.5, 0.8$ 
\item $\Ol=0, 0.1, ...., 1.0$ 
\item $\od=.02, .05, .08, .11, .13, .16, .20, .50$ 
\item $\ob=.003, .015, .018, .020, .022, .025, .03, .04, .07$ 
\item $\ns=0.5, 0.7, 0.8, 0.9, 1.0, 1.1, 1.2, 1.4, 1.7$ 
\item $\nt=-1.0, -0.7, -0.4, -0.2, -0.1, 0$ 
\item $r=0, 0.1, 0.2, 0.3, 0.4, 0.6, 0.8, 1.0, 1.4, 1.8$
\item $\As=0.2,0.3,0.4,0.5,0.6,0.7,0.8,0.9$
\item $b=0.5,0.6,0.7,0.8,0.9,1.0,1.1,1.2,1.3,1.4$ 
\end{itemize}

For each grid point in parameter space, we compute
CMB and matter power spectra using the method detailed in \cite{concordance}. 
We compute the corresponding theoretical lensing predictions
as described in the following section. Although these lensing 
computations follow a fairly standard procedure,  
our procedure is described explicitly below to 
make clear which approximations we make.

\subsection{Lensing calculations}


The lensing measurements that we will use are quoted in terms of the 
variance of the aperture mass $M_{\rm ap}$,
given by\cite{Kaiser96,Schneider98}:
%
\beq{ApermassVariEq}
\expec{M_{\rm ap}^2(\theta)}=2\pi 
\int^{\infty}_0 \ell P_\kappa(\ell)\left[{12\over \pi (\ell\theta)^2}J_4(\ell\theta)\right]^2 d\ell ,
\eeq
where $J_4$ is the fourth-order Bessel function of the first kind
and $P_\kappa(\ell)$ is the
convergence power spectrum, given by \cite{Kaiser96,Schneider98,Jain97,Kaiser92}
\beq{PowerEq}
P_\kappa(\ell) = \frac{9H_0^4\Om^2}{4c^4} 
 \int_0^{w_H} {\overline{W}(w)^2\over a(w)^2} 
 P^{nl}_\delta\left[{\ell\over f_K(w)}, z(w)\right]dw.
\eeq
Here $P^{nl}_\delta(k,z)$ is the nonlinear mass power spectrum, 
and 
$a(w)=[1+z(w)]^{-1}$ is the cosmic scale factor.
$w$ is the radial comoving distance defined by
\beq{ComovingDEq}
w(z) = \int_0^z {c\over H(z')} dz',
\eeq
and $w_H\equiv w(\infty)$ is the comoving distance to the horizon.
The time-variation of the Hubble parameter $H$ is given by
\beq{HubbleEq}
{H\over H_0}=\sqrt{(1+z)^3\Om +(1+z)^2(1-\Om-\Ol)+\Ol}.
\eeq
$f_K(w)$ is the comoving angular diameter distance out to $w$, 
given by
\beq{RadialFEq}
f_K(w) = \left\{\begin{array}{lll}
               K^{-1/2}\sin(K^{-1/2} w) & \mbox{$K>0$} \\
	       w                     & \mbox{$K=0$} \\
	       (-K)^{-1/2}\sinh((-K)^{-1/2} w) & \mbox{$K<0$} 
	       \end{array}
	 \right.
\eeq
where $K$ is the spatial curvature defined by 
\beq{CurvEq}
K\equiv -\left({H_0 \over c}\right)^2\Ok, \quad\Ok\equiv (1-\Om-\Ol).
\eeq
The weighting function $\overline{W}(w)$ is the source-averaged distance
ratio
\beq{WeightFGenEq}
\overline{W}(w)=\int_w^{w_H} G(w'){f_K(w'-w)\over f_K(w')}dw',
\eeq
where $G(w)=G(w(z))$ is the source redshift distance distribution,
usually approximated by a fitting function of the form
\beq{SourceDistEq}
G(w) = {\beta\over z_0 \Gamma ({1+\alpha\over \beta})} 
\left({z\over z_0}\right)^{\alpha}e^{-(z/z_0)^{\beta}},         
\eeq
with fitting parameters $z_0$,$\alpha$ and $\beta$.

The linear power spectrum at
redshift $z$ is related to the present one by
\beq{linearzEq}
P^{\rm linear}_\delta(z,k)= \left[{g(z)\over g(0)}\right]^2 P^{\rm linear}_\delta(0,k) ,
\eeq
where the linear growth factor $g(z)$ is approximately given by \cite{Carroll92} 
\def\Omz{\Om^z}
\def\Olz{\Ol^z}
\beq{GrowthFZEq}
g(z)\approx {5\over 2}\Omz
 \left[{\Omz}^{4/7}-\Olz+\left(1+{\Omz\over 2}\right)\left(1+{\Olz\over 70}\right)\right]^{-1}.
\eeq
Here the parameters $\Omz$ and $\Olz$ are those at redshift $z$, 
given by the present values $\Om$ and $\Ol$ through the relations 
\beq{OmEq}
\Omz= \left[{H_0\over H(z)}\right]^2 \Om (1+z)^3,
\eeq
\beq{OlEq}
\Olz = \left[{H_0\over H(z)}\right]^2 \Ol,
\eeq
with $H(z)$ given by \eq{HubbleEq}.

The power spectrum $P^{nl}_\delta(k)$ needed in \eq{PowerEq} is the
nonlinear one rather than the linear one $P^{l}_\delta(k)$ 
given by \eq{linearzEq}.
Based on a pioneering idea of Hamilton {\etal} \cite{Hamilton91},
a series of approximations 
\cite{Jain97,Jain95,PeacockDodds94,PeacockDodds96} 
have been developed for approximating the former using the latter.
In terms of the dimensionless power  
\beq{DlessPEq}
\Delta^2(k) \equiv {4\pi\over (2\pi)^3}k^3 P_\delta(k),
\eeq
the linear power $\Deltal$ on scale $\kl$ is approximately related to the nonlinear power
$\Deltanl$ on a smaller nonlinear scale $\knl$.
We use the Peacock \& Dodds' approximation \cite{PeacockDodds96}, 
where this mapping is given by 
\beq{LtoNlEq}
\Delta_{nl}^2(k_{nl})=f_{nl}\left[\Delta_l^2(k_l)\right]
\eeq
and
\beq{LktoNlkEq}
k_l=\left[1+\Delta_{nl}^2(k_{nl})\right]^{-1/3} k_{nl},
\eeq
with a fitting function 
\beq{FittingEq}
f_{nl} (x)=x \left\{ {1+B\beta x+(Ax)^{\alpha \beta} \over 
1+\left[{(Ax)^{\alpha} g(0)^3 \over (Vx^{1/2})}\right]^{\beta}}
\right\}^{1/\beta},
\eeq
parametrized by 
\beqa{PforFittingEq}
A     & = & 0.482(1+\neff/3)^{-0.947},\\
B     & = & 0.226(1+\neff/3)^{-1.778},\\
\alpha& = & 3.310(1+\neff/3)^{-0.224},\\
\beta & = & 0.862(1+\neff/3)^{-0.287},\\
V     & = & 11.55(1+\neff/3)^{-0.423}.
\eeqa
Here $g(0)$ is the linear growth factor of \eq{GrowthFZEq} evaluated at $z=0$ and
$\neff\equiv d\ln\Pl/ d\ln\kl$ is the effective logarithmic slope of the 
linear power spectrum evaluated at $\kl$. Since this slope should be evaluated
for a model without baryonic wiggles, we compute $\neff$ using an
the Eisenstein \& Hu fitting function with baryon oscillations turned off.


\section{Experimental data used}
\label{DataSec}

\subsection{CMB data}
\label{CMBdata}

\begin{figure}[tb] 
\centerline{\epsfxsize=9.0cm\epsffile{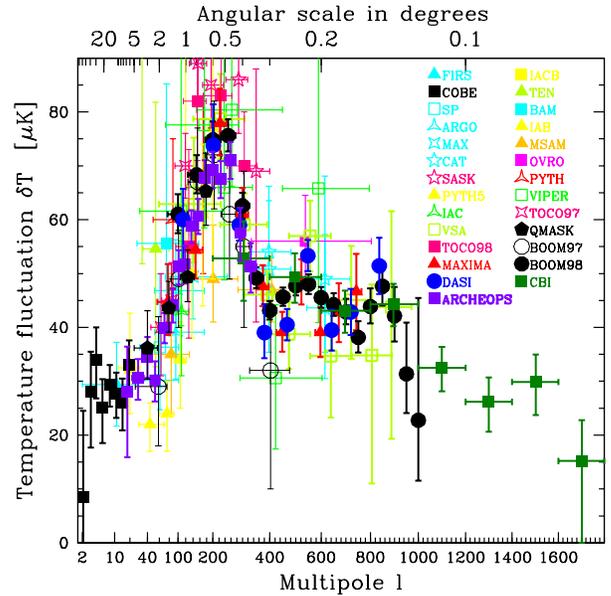}}
\vskip-0.8cm
\smallskip
\caption{\label{cmbdataFig}\footnotesize%
CMB data used in our analysis.
Error bars do not include calibration or beam errors
which allow substantial vertical shifting and 
tilting for some experiments (these effects are included in our analysis).
}
\end{figure}

\Fig{cmbdataFig} shows the 135 CMB measurements which are used in our
analysis. Compared to the data set we used in \cite{xiaomin01}, we add the new
measurements from the Cosmic Background Imager (CBI) mosaic\cite{CBI02}, 
the Very Small Array (VSA) \cite{VSA02c} and Archeops \cite{Archeops02}.
For CBI, we use the year 2000 observations of 
three pairs of mosaic fields \cite{CBI02} but not 
the deep fields, because it is still unclear whether their signal is dominated by CMB 
or other effects such as SZ effect\cite{CBI02b}.
The Boomerang results updated last week \cite{Ruhl02} and the Acbar results \cite{Acbar02}
became available too recently for inclusion in this analysis, but 
we do include them in the online combined power spectrum described below.

%

\begin{figure}[tb] 
\centerline{\epsfxsize=9.0cm\epsffile{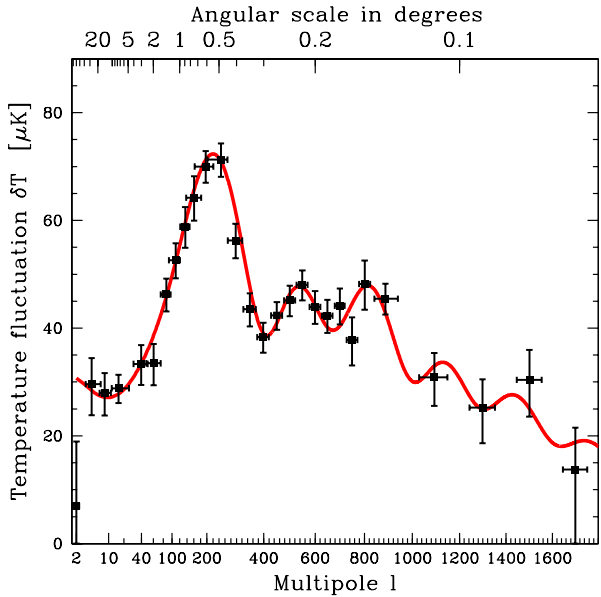}}
\vskip-0.8cm
\smallskip
\caption{\label{cmb_comboFig}\footnotesize%
Combination of data from \protect\fig{cmbdataFig}.
These error bars include the effects of beam and
calibration uncertainties, which cause long-range correlations
of order 10\% over the peaks. In addition, points tend to be anti-correlated 
with their nearest neighbors, typically at the level of 10-20\%.
The curve shows our model best fitting CMB+LSS data (second last column in Table 2).
}
\end{figure}

We combine these measurements into a single set of 28 band powers shown in
\fig{cmb_comboFig} and Table 1 using
the method of \cite{xiaomin01} as improved in \cite{Tegmark02}, including calibration and beam 
uncertainties, which effectively calibrates the experiments against each other. 
Since our compressed band powers $d_\l$ are simply linear combinations 
of the original measurements, they can be
analyzed ignoring the details of how they were constructed, being 
completely characterized by a window 
matrix $\W$:
\beq{lwindowEq}
\expec{d_i} = \sum_{\l}\W_{i\l}\delta T_\l^2,
\eeq
where $\delta T_\l^2\equiv\l(\l+1)C_\l/2\pi$  
is the angular power spectrum.
This matrix is available at $www.hep.upenn.edu/{\sim}max/cmb/cmblsslens.html$
together with the 28 band powers $d_\l$ and their 
$28\times 28$ covariance matrix.
The data $\l$-values and effective $\l$-ranges in \fig{cmb_comboFig} and Table 1
correspond to the median, 20th and 80th percentile of the window functions $\W$.
Comparing Table 1 with the older results from \cite{Tegmark02}, we find that
the only major change is a shallower rise towards the 1st peak
due to Archeops, which is able to help calibrate Boomerang and other 
small-scale experiments by connecting them with the COBE.
Specifically, $\delta T_\l$ has increased by
about 10\% at $\l\sim 50$ and 
decreased about 5\% for $\l\sim 100-200$, thereby nudging the first peak a tad to the right.

\bigskip
\noindent
{\footnotesize
{\bf Table 1} -- Band powers combining the
information from CMB data 
from \protect\fig{cmbdataFig}.
The 1st column gives the $\l$-bins used when combining the data, and can be ignored
when interpreting the results.
The 2nd column gives the medians and characteristic widths of 
the window functions as detailed in the text. 
The error bars in the 3rd column
include the effects of calibration and beam uncertainty.
The full $28\times 28$ correlation matrix and $28\times 2000$
window matrix are available at $www.hep.upenn.edu/\sim max/cmb/cmblsslens.html$.
\bigskip
\begin{center}
{\footnotesize
\begin{tabular}{|c|c|c|}
\hline
$\l$-Band	&$\l$-window	&$\delta T^2\>[\mu $K$^2]$\\
\hline		
$    2-   2$	 &$    2_{- 0}^{+ 0}$&$  49\pm 310$\\
$    3-   5$	 &$    4_{- 1}^{+ 3}$&$ 877\pm 308$\\
$    6-  10$	 &$    8_{- 2}^{+ 3}$&$ 782\pm 218$\\
$   11-  30$	 &$   16_{- 4}^{+ 9}$&$ 832\pm 151$\\
$   31-  50$	 &$   40_{-10}^{+10}$&$1113\pm 244$\\
$   51-  75$	 &$   60_{-13}^{+14}$&$1120\pm 255$\\
$   76- 100$	 &$   87_{-12}^{+10}$&$2139\pm 279$\\
$  101- 125$	 &$  110_{-17}^{+11}$&$2767\pm 340$\\
$  126- 150$	 &$  135_{-14}^{+12}$&$3461\pm 443$\\
$  151- 175$	 &$  161_{-23}^{+21}$&$4122\pm 529$\\
$  176- 225$	 &$  196_{-34}^{+24}$&$4900\pm 410$\\
$  226- 275$	 &$  246_{-44}^{+23}$&$5079\pm 441$\\
$  276- 325$	 &$  297_{-28}^{+24}$&$3164\pm 359$\\
$  326- 375$	 &$  348_{-23}^{+22}$&$1892\pm 265$\\
$  376- 425$	 &$  398_{-22}^{+20}$&$1468\pm 213$\\
$  426- 475$	 &$  450_{-22}^{+21}$&$1793\pm 219$\\
$  476- 525$	 &$  499_{-22}^{+21}$&$2037\pm 257$\\
$  526- 575$	 &$  549_{-23}^{+21}$&$2306\pm 268$\\
$  576- 625$	 &$  600_{-22}^{+21}$&$1932\pm 267$\\
$  626- 675$	 &$  649_{-22}^{+21}$&$1790\pm 259$\\
$  676- 725$	 &$  700_{-21}^{+20}$&$1948\pm 293$\\
$  726- 775$	 &$  749_{-23}^{+22}$&$1428\pm 334$\\
$  776- 825$	 &$  801_{-23}^{+23}$&$2322\pm 438$\\
$  826-1000$	 &$  888_{-46}^{+52}$&$2067\pm 261$\\
$ 1001-1200$	 &$ 1093_{-65}^{+56}$&$ 953\pm 300$\\
$ 1201-1400$	 &$ 1299_{-55}^{+54}$&$ 638\pm 291$\\
$ 1401-1600$	 &$ 1501_{-55}^{+54}$&$ 924\pm 368$\\
$ 1601-\infty$   &$ 1700_{-53}^{+51}$&$ 189\pm 273$\\
\hline		
\end{tabular}
}
\end{center}
}


\subsection{LSS data}
\label{LSSdata}

Measurements of $P(k)$ from Galaxy redshift surveys have recently improved in 
both quality and quantity, 
and the Sloan Digital Sky Survey is set to continue this trend.
In this paper, we use the power
spectrum from the 2dFGRS\cite{Colless01} as measured by \cite{Tegmark2df01}.
We model the galaxy bias as a scale-independent constant $b$, and therefore
discard all 2dF measurements with $k\ge 0.3h/$Mpc to minimize our sensitivity
to nonlinear clustering and nonlinear bias effects.

\begin{figure}[tb] 
\vskip1.0cm
\centerline{\epsfxsize=9.0cm\epsffile{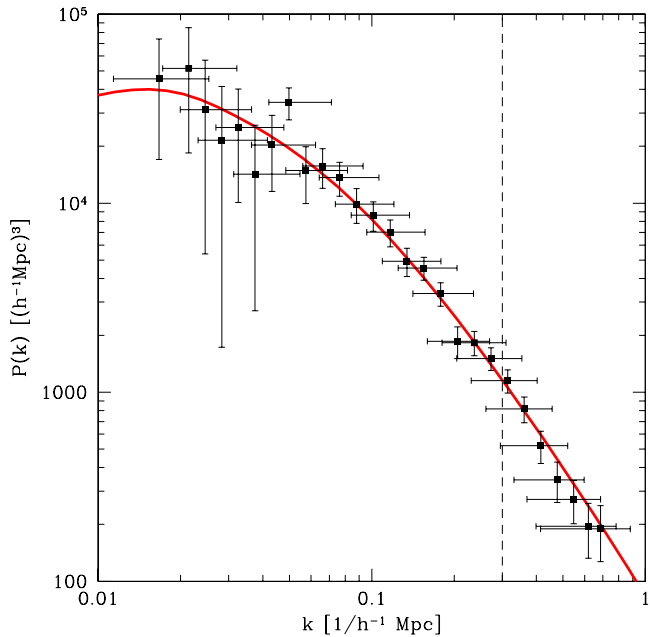}}
\vskip0.8cm
\smallskip
\caption{\label{lssdataFig}\footnotesize%
2dF galaxy power spectrum data used.
We include only points leftward of the dotted line (with $k<0.3h/Mpc$) in our analysis
to stay in the linear regime.
The curve shows our model best fitting CMB+LSS data (second last column in Table 2).
}
\end{figure}

\subsection{Lensing data}
\label{Lensingdata}

For this paper, we use the results from Hoekstra \etal\cite{Hoekstra02a}, 
the analysis of 53 square degrees of $R_C$-band imaging data
from the Red Sequence Cluster Survey(RCS). The RCS\cite{RCS01a,RCS01b} 
is a 100 square degree galaxy
cluster optical survey designed to provide a large sample of 
optically selected clusters
of galaxies with redshifts $0.1<z<1.4$
We use only the seven data points which have been used in cosmological
analysis in\cite{Hoekstra02a}. 
We model the source redshift distribution, we use the multi-redshift part in 
\eq{SourceDistEq}, with $\alpha=4.7$, $\beta=1.7$ and $z_0=0.302$, which are
the best fit value given by Hoekstra \etal\cite{Hoekstra02a}.

\begin{figure}[tb] 
\vskip1.0cm
\centerline{\epsfxsize=9.0cm\epsffile{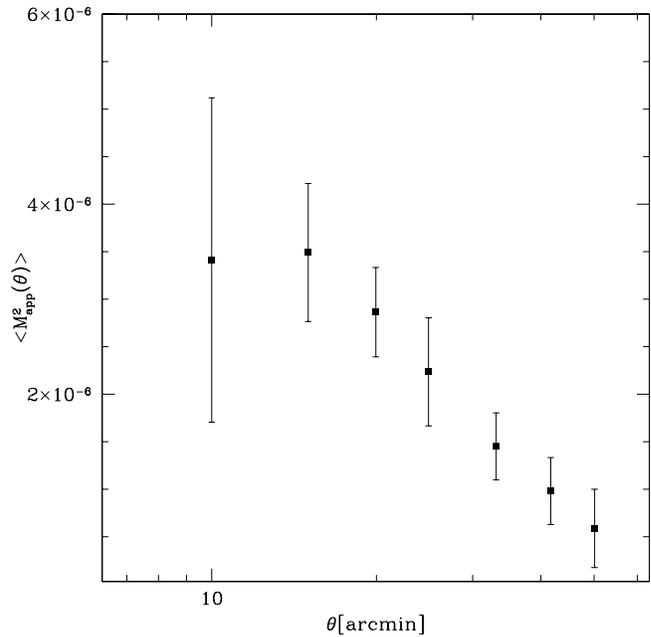}}
\vskip0.8cm
\smallskip
\caption{\label{lensingdataFig}\footnotesize%
Lensing data used in our analysis.
The two curves show our best fit model excluding (bottom) and including (top) lensing data.
}
\end{figure}

\section{Parameter constraints}
\label{ParameterSec}

Our cosmological parameter constraints are summarized in Table 2.
We study three cases: using CMB data only, adding LSS information and adding 
lensing  measurements too, respectively. These three cases correspond 
to the three columns of the table and are discussed in the corresponding 
three subsections below.

\subsection{From CMB data alone}
\label{From CMB data alone}

\bigskip

\begin{figure}[tb] 
\centerline{\epsfxsize=8.5cm\epsffile{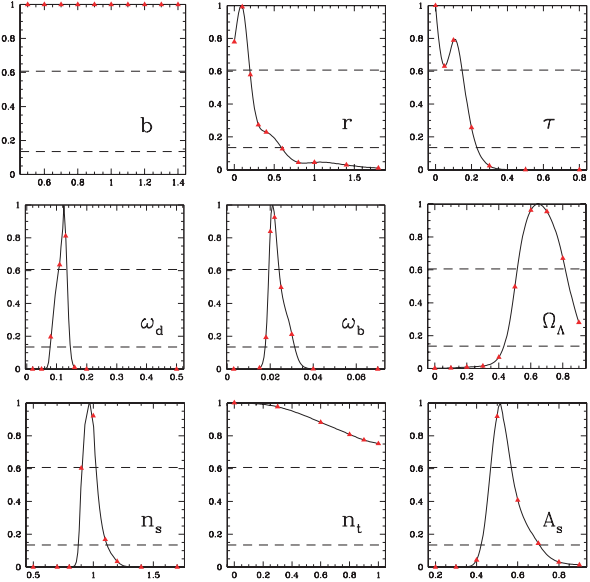}}
\bigskip
\caption{\label{CMB1DnoFig}\footnotesize%
Constraints on individual parameters using only CMB information.
The quoted $1-\sigma$ and $2-\sigma$ confidence limits are where each 
curve drops below the horizontal dashed lines $e^{-1^2/2}\approx 0.61$ and 
$e^{-2^2/2}\approx 0.14$, respectively.
}
\end{figure}

\Fig{CMB1DnoFig} shows our constraints on nine individual cosmological parameters,
using only the CMB data.
Comparing to numerous previous analyses using CMB data alone \cite{10par,concordance,Efstathiou01,xiaomin01}, 
we see that the addition of new CMB data from CBI, Archeops {\etc} has 
substantially tightened the constraints, especially on the tensor-to-scalar ratio $r$ 
and dark energy $\Ol$.
CMB alone is seen to firmly rules out models with no dark matter 
and models with no dark energy.
(The dark energy constraint depends critically on our prior assumption $\Ok=0$,
whereas the dark matter result does not.)

The last four constraints in Table 2 are computed with the moments method as in \cite{xiaomin01}.
They are for parameters ($h$, $\beta$, $\zion$ and $\sigma_8$)
that are not independent and merely functions of the 11 parameters that
define our model grid.
The Hubble parameter $h$ is given by 
\beq{hEq}
h = \sqrt{\od+\ob\over 1-\Ok-\Ol}.
\eeq
The redshift-space distortion parameter $\beta$ is 
given by \cite{Hamilton00,PeeblesBook80,Lahav91}
\beq{betaEq}
\beta={f(\Om,\Ol)\over b},
\eeq
where $f(\Om ,\Ol)\approx \Om ^{\rm 0.6}$ is the velocity-suppression factor\cite{PeeblesBook80,Lahav91}
(we evaluate $f$ exactly using the grow$\lambda$ package \cite{Hamilton00}).
If the diffuse intergalactic hydrogen was reionized abruptly at a redshift $\zion$, then
\cite{PeeblesBook93}.
\beq{ReionZEq}
z_{\rm ion}\approx 8.9\left({{\tau h}\over\ob}\right)^{2/3} \Om^{1/3}.
\eeq
in the approximation that $\zion\gg 1$ (which we know to be the case).
The alternative normalization parameter $\sigma_8$ is the rms fluctuation in an $8h^{-1}$ sphere,
given by a weighted integral of the matter power spectrum \cite{PeeblesBook80}.

\bigskip
\bigskip
\noindent
{\footnotesize
{\bf Table 2} -- Best fit values and $1-\sigma$ confidence limits on
cosmological parameters.
For the numbers above the horizontal line, 
the central values are the ones maximizing the likelihood (the best fit
model).
For the numbers below the horizontal line, 
the central values are
means rather than those for the best fit model\protect\cite{xiaomin01}.
For instance, the Hubble parameters for the best fit models
are $h=0.68$, $0.69$ and $0.61$ for the three columns, respectively, 
which differs from the mean values in the table.
\def\fnp{{$\sim$0}}
\bigskip
\begin{center}
{
\begin{tabular}{|l|c|c|c|}
\hline
              &CMB alone              &$+$ 2dF                 &$+$ RCS
lensing          \\
\hline
$\tau$                &\zzp{0.04}{.06}        &\zz{0.06}{.03}{.03}
&\zz{0.20}{.03}{.03}      \\
$\Ol$         &\zz{0.71}{.11}{.11}    &\zz{0.72}{.09}{.09}
&\zz{0.59}{.08}{.08}      \\
$h^2\Od$      &\zz{0.112}{.014}{.014}    &\zz{0.115}{.013}{.013}
&\zz{0.128}{.008}{.008}      \\
$h^2\Ob$      &\zz{0.023}{.003}{.003} &\zz{0.024}{.003}{.003}
&\zz{0.024}{.002}{.002}   \\
$n_s$         &\zz{0.99}{.06}{.06}    &\zz{0.99}{.04}{.04}
&\zz{1.00}{.02}{.02}      \\
$As$            &\zz{0.54}{.07}{.07}    &\zz{0.55}{.06}{.06}
&\zz{0.78}{.05}{.05}      \\
$r$             &\zzp{0.10}{.17}        &\zzp{0.08}{.11}
&\zzp{0.02}{.05}          \\
$b$             &                       &\zz{0.87}{.06}{.06}
&\zz{0.89}{.04}{.04}      \\
\hline        
$h$           &\zz{0.71}{.13}{.13}    &\zz{0.73}{.11}{.11}
&\zz{0.61}{.04}{.04}     \\
$\beta$               &\zz{0.40}{.16}{.16}    &\zz{0.39}{.07}{.07}
&\zz{0.46}{.04}{.04}  \\
$\zion$               &\zzp{4.82}{5,97}      &\zz{8.06}{2.04}{2.04}
&\zz{19.23}{2.45}{2.45}   \\
$\sigma_8$      &\zz{0.72}{.06}{.06}    &\zz{0.74}{.06}{.06}
&\zz{0.92}{.04}{.04} \\
\hline                
$\chi^2/n$	&$23.7/21$		&$38.7/40$		&$61.2/47$\\		
\hline                
\end{tabular}
}
\end{center}
}


\subsection{From CMB+LSS data}
\label{From CMB+LSS data}

\bigskip

\begin{figure}[tb] 
\centerline{\epsfxsize=8.5cm\epsffile{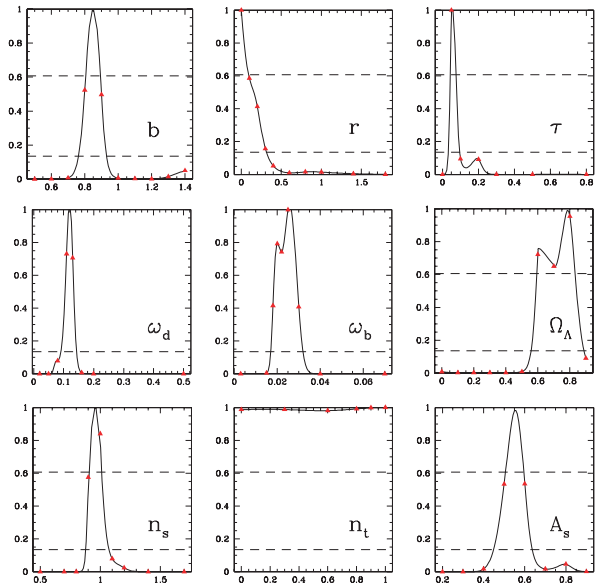}}
\bigskip
\caption{\label{CMBLSS1DnoFig}\footnotesize%
Same as previous figure but combining CMB and LSS information.
}
\end{figure}

\Fig{CMBLSS1DnoFig} shows how the above-mentioned constraints tighten 
up when including the 2dF galaxy clustering data. The improvement comes 
mainly from breaking degeneracies involving 
curvature, dark energy, dark matter, baryons and tilt.
The $\Ol$-measurement is seen to be nicely consistent with the
many other pieces of evidence (SN Ia {\etc}) pointing towards $\Om\approx 0.7)$.
The 2dF bias value $b\approx 0.87\pm 0.06$ (interpreted as the real-space bias at the 
effective survey redshift) is consistent with that found by the 2dF 
team \cite{Lahav01,Verde02}, although on the low side (since we are marginalizing
 over $b$, this has no impact on our constraints on other parameters such as $\sigma_8$).  
The most striking improvement is seen to be for the reionization optical depth $\tau$, 
hinting at a detection of reionization around $z\approx 8$.
However, as we will discuss at length in \sec{DiscussionSec}, the true significance
of this detection is probably lower than \fig{CMBLSS1DnoFig} suggests.

\subsection{From CMB, LSS and lensing data}
\label{From CMB, LSS and lensing data}

\subsubsection{One-Dimensional Constraints}

\bigskip

\begin{figure}[tb] 
\centerline{\epsfxsize=8.5cm\epsffile{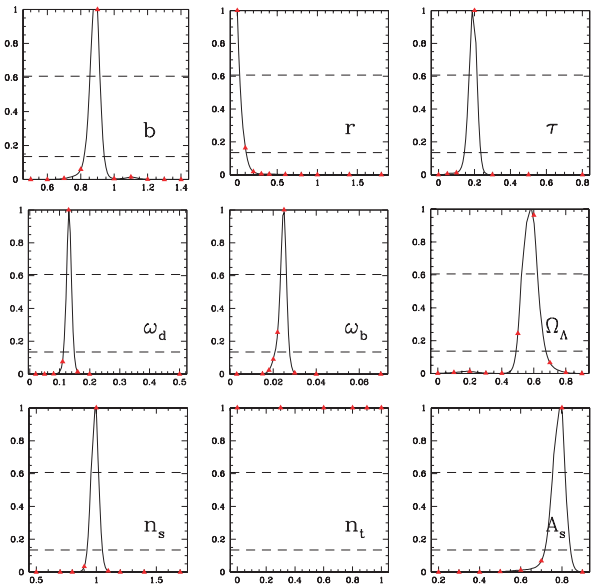}}
\bigskip
\caption{\label{CMBLSSLENS1DnoFig}\footnotesize%
Same as previous figure but combining CMB, LSS and lensing information.
}
\end{figure}

Above we saw that adding 2dF data to the CMB not only tightened parameter constraints, but also 
produced a consistent picture that agreed well with other measurements supporting the 
cosmological concordance model.
\Fig{CMBLSSLENS1DnoFig} shows that things are quite different when adding the RCS lensing data.
Although the nominal constraints do get substantially sharper as predicted \cite{Hu98}, 
most strikingly for the  
tensor-to-scalar ratio $r$, the harmonious consistency becomes clouded.
Adding the 7 lensing data points raises $\chi^2$ by $22.5$ (from 38.7 to 61.2 for
about 54 effective degrees of freedom\footnote{
The effective number of degrees of freedom is the number of of data points
minus the effective number of parameters fit for.
Although we nominally used 9 parameters, $b$ has an effect only when LSS is 
included and $\nt$ has no effect in the absence of tensors (and $r$ is indeed consistent with zero).
The effective number of degrees of freedom is therefore about
$28-7=21$ for the CMB-only case, 
$28+20-8=40$ for the CMB+LSS case and 
$28+20+7-8=47$ for the CMB+LSS+lensing case.
}). Although this per se might be considered
acceptable, it is striking how much the lensing shifts the best-fit values
of many parameters.
Most notably, the reionization optical depth $\tau$ increases from $0.06\pm0.03$ 
to $0.20\pm 0.03$\footnote{
The Gunn-Peterson constraint implies that our Universe is highly ionized for $z\simlt 6$,
and the recent detection of a Gunn-Peterson trough in $z\sim 6$
quasars \cite{Becker01} has been interpreted as the tail end of reionization.
Moreover, it has been argued that the temperature of the intergalactic medium at
$z\approx 3.4$ inferred from He II reionization requires $z<9$ \cite{Theuns02a,Theuns02b}, nicely
consistent with our best fit CMB+LSS value $z\approx 8$.
However, models will less abrupt reionization still allow substantially 
larger $\tau$-values consistent with our CMB+LSS+RCS 
constraint $\tau=0.20\pm 0.03$ \cite{Venkatesan02,Kaplinghat02,WyitheLoeb02},
so the issue of whether to trust the lensing normalization cannot yet be settled by
reionization constraints.
} 
and the best fit cosmological constant $\Ol$ drops
from $0.71$ to $0.55$, which sits uncomfortably with, \eg, SN Ia constraints.
The cosmological model is 
pushed into an uncomfortably configuration by the lensing data,
creaking and squeaking but refusing to outright collapse.

As we will see in \sec{DiscussionSec}, this tension comes from the overall normalization 
of the power spectrum, with the RCS lensing data pushing the best-fit $\sigma_8$ value
up from $0.71$ to $0.91$. This normalization agrees with that of some
other cosmic shear surveys \cite{Bacon02,Hoekstra02a,Refregier02,Waerbeke02}, 
but not with that preferred by CMB and LSS data. In \sec{DiscussionSec}, 
we will comment on uncertainties and other lensing data with different normalization.


\subsubsection{Two-Dimensional Constraints}
For more detailed insight into our parameter constraints and degeneracies, 
we now turn to joint constraints on pairs of parameters.

\begin{figure}[tb] 
\centerline{\epsfxsize=8.5cm\epsffile{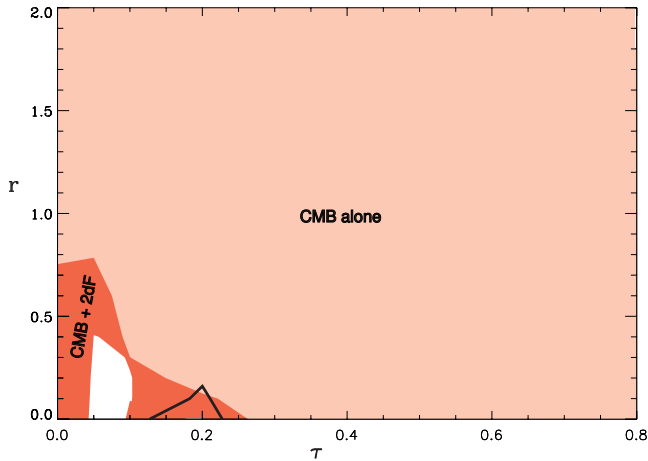}}
\bigskip
\caption{\label{tarFig}\footnotesize%
Constraints in the $(\tau,r)$-plane. 
The large pink/light grey region is ruled out by CMB alone at the 95\% confidence
level. The red/grey area shows the region excluded by adding additional constraints from 2dF. 
The black curve shows
the allowed region when combine CMB, 2dF and RCS lensing.
}
\end{figure}

\begin{figure}[tb] 
\centerline{\epsfxsize=8.5cm\epsffile{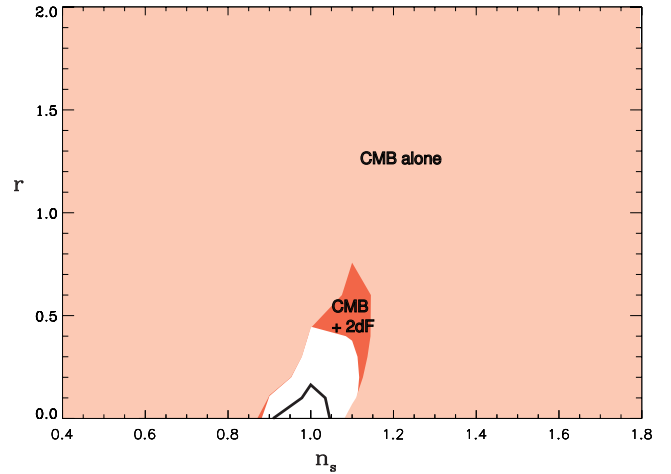}}
\bigskip
\caption{\label{nsrFig}\footnotesize%
Constraints in the $(\ns,r)$ plane.
The large pink/light grey region is ruled out by CMB alone. 
The red/grey region becomes ruled out when adding 2dF information. 
The black curve shows the allowed region when combine CMB, 2dF
and RCS lensing.
}
\end{figure}

\Fig{tarFig} shows that 2dF breaks the CMB degeneracy between reionization and gravity waves
and produces essentially independent (separable) constraints on $\tau$ and $r$.
The gravity wave limits agree well with those found by \cite{Melch02c} who also included 2dF and Archeops data.
The fact that the allowed regions excluding and including lensing data have essentially
no overlap illustrates the poor consistency between CMB+LSS and lensing data sets mentioned above.

\Fig{nsrFig} shows constraints in the parameter space $(\ns,r)$ where inflationary
models can be tested. 
These limits from CMB only and from CMB+LSS are comparable with those 
in older work requiring stronger priors.
For instance, comparison with a corresponding plot from a year ago \cite{xiaomin01} 
shows new data has squeezed the constraints substantially, particularly on the gravity
wave amplitude.
\footnote{
As discussed in \cite{Martin00}, there has been a fair amount of
notational confusion in the literature surrounding the 
tensor-to-scalar ratio $r$.
There are two logical ways to define this ratio:
either in terms
of the fundamental parameters of the power spectrum (or, 
equivalently, of the inflationary model space),
or in terms of the observables, usually the CMB quadrupoles.
As in \cite{xiaomin01}, we adopt the former approach and define 
\beq{RDefEq}
r\equiv {A_t\over A_s},
\eeq
where $A_s$ and $A_t$ are the scalar and tensor fluctuation amplitudes
as defined in \cite{Martin00}.
For inflation models where the slow-roll approximation is valid, 
this ratio is related to the tensor tilt $\nt$ by the so-called
inflationary consistency condition \cite{Martin00,Liddle92}
\beq{InfConsistencyEq}
r = -{200\over 9} \nt.
\eeq
A common alternative definition of the tensor-to-scalar ratio is
the quadrupole ratio
\beq{rDefEq}
R \equiv {C^{\rm tensor}_2\over C^{\rm scalar}_2},
\eeq
Writing the ratio $r/R$ is typically between 2 and 5 --- it depends on the values of $\Ol$ and $\Ok$ 
via the late integrated Sachs-Wolfe effect. 
}
Now CMB data alone gives $r<0.48$ at 95\% level, obtainable only be combining 
CMB and LSS data back then. Now CMB+2dF lowers the limit to $r<0.34$.
The lensing data nominally helps even helps more, tightening the constraint to $r<0.13$ --- we return
to the issue of whether this is believable in \sec{DiscussionSec}.

\begin{figure}[tb] 
\centerline{\epsfxsize=8.5cm\epsffile{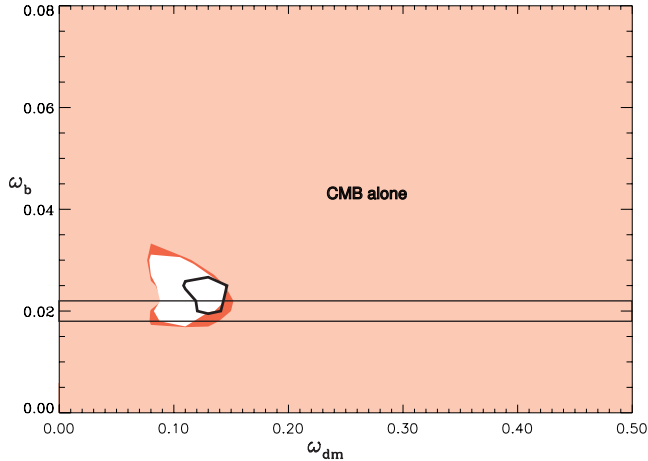}}
\bigskip
\caption{\label{odobFig}\footnotesize%
Constraints in the $(\od,\ob)$-plane. 
The large pink/grey region is ruled out by CMB data alone at
95\% confidence. The smaller red/shaded region becomes excluded
when imposing additional constraints from 2dF. 
The thick black curve shows the allowed region when combine CMB, 2dF
and RCS. The rectangular band by the thin straight lines shows the BBN constraints from \protect\cite{Burles01}.
}
\end{figure}

Let us now turn to the cosmic matter budget.
A year ago \cite{xiaomin01,Efstathiou01}, 
the CMB+LSS constraints on the baryon density $\ob$ were beautifully consistent with 
those from Big Bang Nucleosynthesis (BBN), $\ob=0.02\pm0.002$.
Although the agreement is still acceptable, \fig{odobFig} shows that the CMB+LSS baryon density
is creeping back up again towards the higher range that was favored two years
ago \cite{Lange00,boompa}.
This slight tension is reflected in $\chi^2$ increasing by 3 if we add a the BBN prior to our
likelihood analysis.

\section{Discussion}
\label{DiscussionSec}

Since cosmic shear measurements have now improved to the point where they deserve
to be treated on par with CMB and galaxy clustering data, we have 
perform a detailed 9-parameter analysis of recent lensing, 
CMB and galaxy clustering data.

For the CMB part, we constructed and used a data set combining all available 
data taking calibration errors and beam uncertainties into account, effectively 
calibrating them off of each other. 
Using this combined power spectrum for a ``MAP versus world'' comparison
next month will provide a powerful test of how realistic the error estimates have been 
in the CMB community.

Even though 2002 has added substantial amounts of information on the CMB power spectrum,
\eg, from CBI and Archeops, the combined CMB and 2dF data remains comfortably
consistent with a simple flat adiabatic scale-invariant concordance
model with $\Ol=0.72\pm 0.09$, $h^2\Oc=0.115\pm 0.013$ and
$h^2\Ob=0.024\pm 0.003$. The new data has added a 
hint of reionization around $z\sim 8$. These constraints assumed negligible contributions 
from spatial curvature and massive neutrinos.

Including the RCS lensing data in our analysis 
confirmed with real data what was predicted by Fisher matrix forecasts
\cite{Hu98}: that
lensing has the statistical power to substantially help with cosmological parameter constraints
in two ways: by breaking degeneracies and by providing cross-checks.
Intriguingly, one of the cross-checks failed, so let us now turn to this failure in more detail.

\subsection{The $\sigma_8$ problem}

Our results suggest that the tension between the lensing and other data comes from 
the overall normalization of the power spectrum. 
We saw that the CMB+2dF data preferred $\sigma_8=0.74$, in good agreement 
with results from last year \cite{Lahav01}. In contrast, the
RCS lensing data prefers higher values \cite{Hoekstra02a} and pushed the normalization up 
to $\sigma_8=0.92$ in our joint analysis.

Physically, it is easy to see why this pushed other parameters in the way it did, notably
increasing $\tau$ and decreasing $\Ol$.
If we increase the value of $\sigma_8$, the CMB and galaxy power spectra will both increase
in amplitude. This causes no problems for fitting the 2dF data, since our method can merely
readjust their bias, but causes serious difficulties with the CMB.
The acoustic peaks can be brought back down to acceptable heights by increasing $\tau$,
but this provides only a partial solution to the problem, since it leaves the power on very
large $\l\simlt 10$ scales larger than observed by COBE DMR.
Lowering $\Ol$ helps with this, since it increases the linear growth factor and thereby 
lowers the entire CMB power spectrum for fixed current matter clustering.
$\Ol$ can only be lowered by a moderate amount, however, since
the matter density $\Om\sim 1-\Ol$ will increase and give the galaxy power spectrum the wrong shape.
The response of the parameter fitting algorithm is therefore to do a little bit of both
tweaks, increasing $\tau$ and lowering $\Ol$.
Indeed, similar results were obtained in a smalled parameter space by Hoekstra {\etal} \cite{Hoekstra02a},
who found that they could only reconcile their lensing normalization with CMB for a rather
large reionization optical depth $\tau=0.12\pm0.04$.

To verify that this is the correct physical interpretation of what is going on, 
we repeated our entire analysis with the amplitude of the 
RCS data lowered by a factor of $(0.7/0.86)^2$, 
which is roughly the ratio of normalizations preferred by CMB+2dF and RCS respectively.
This analysis gave results fully consistent with those from CMB+2dF alone, 
for example, 
$\tau=0.05\pm0.03$, $\Ol=0.76\pm0.13$, $\od=0.11\pm0.02$, $\ob=0.024\pm0.005$,
$\ns=1.00\pm0.04$, $\As=0.52\pm0.10$, $r=0.10+0.23$, $b=0.82\pm0.09$, $z_{ion}=7.64\pm3.18$ 
and $\sigma_8=0.70\pm0.11$. These results not only demonstrate consistency, 
but also show that by combining lensing and CMB and LSS, 
we can get stronger constraints on parameters like 
$\Ol$, $\ns$, $\sigma_8$ and $z_{ion}$, just as had been predicted using information theory \cite{Hu98}. 

\def\sss{\hglue5mm}
\bigskip
\noindent
{\footnotesize
{\bf Table 3} -- Recent measurements of the power spectrum normalization $\sigma_8$.
For results quoted as fitting formulas involving $\Om$, we have set $\Om=0.3$.
Error bars are $1\sigma$ --- for \cite{Hoekstra02a} and \cite{Jarvis02}, 
we have simply divided their quoted 95\% errors by 2.
\bigskip
\begin{center}
{\footnotesize
\begin{tabular}{lll}
\hline
Analysis&&$\sigma_8$\\
\hline
{\bf Clusters:}\\
\sss Pierpaoli {\etal} (2001)        &\cite{Pierpaoli01}     &$1.02^{+0.07}_{-0.08}$\\
\sss Borgani {\etal} (2001)          &\cite{Borgani01}       &$0.76^{+0.08}_{-0.05}$\\
\sss Reiprich \& B\"ohringer (2001)  &\cite{Reiprich01}      &$0.68^{+0.08}_{-0.06}$\\
\sss Seljak {\etal} (2001)           &\cite{Seljak01}        &$0.75\pm 0.06$\\
\sss Viana {\etal} (2001)            &\cite{Viana01}         &$0.61\pm 0.05$\\
\sss Bahcall {\etal} (2002)          &\cite{Bahcall02}       &$0.72\pm 0.06$\\
\sss Pierpaoli {\etal} (2002)        &\cite{Pierpaoli02}     &$0.77^{+0.05}_{-0.04}$\\
\hline
{\bf Weak lensing:}\\
\sss Jarvis {\etal} (2002)           &\cite{Jarvis02}        &$0.71^{+0.06}_{-0.08}$\\
\sss Brown {\etal} (2002)            &\cite{Brown02}         &$0.74\pm 0.09$\\
\sss Hoekstra {\etal} (2002)         &\cite{Hoekstra02a}     &$0.86^{+0.04}_{-0.05}$\\
\sss Van Waerbeke{\etal} (2002)      &\cite{Waerbeke02}      &$0.97\pm 0.06$\\
\sss Bacon{\etal} (2002)             &\cite{Bacon02}         &$0.97\pm 0.13$\\
\sss Refregier{\etal} (2002)         &\cite{Refregier02}     &$0.94\pm 0.14$\\
\hline
{\bf CMB:}\\
\sss Lahav {\etal} 2001            &\cite{Lahav01}         &$0.73\pm 0.05$\\
\sss Melchiorri \& Silk 2002       &\cite{Melch02a}        &$0.70\pm 0.05$\\
\sss Lewis {\etal} 2002            &\cite{Lewis02}         &$0.79\pm 0.06$\\
\sss This work           	     &                       &$0.74\pm 0.06$\\
\hline		
\end{tabular}
}
\end{center}
}

So is the CMB normalization too low or is the lensing normalization too high?
As shown in Table 3, $\sigma_8$ is emerging as the currently most controversial cosmological 
parameter, and it will be crucial to use further cosmological 
data to get to the bottom of this issue.
The table illustrates what has been emphasized by many authors, namely that the
scatter between $\sigma_8$ measurements using different techniques is substantially
larger than the formal error bars, triggering unpleasant flashbacks to the 
bimodal and bitter debate about the Hubble parameter $h$.
Although much work clearly needs to be done to resolve this issue, there are some 
encouraging indications that the $\sigma_8$ gap may be closing.
The CMB normalization has increased slightly in the last year, with the 
first acoustic peak growing by $10\%$ 
between \cite{xiaomin01} and our present compilation.
CMB experiments can now be more effectively calibrated off of each other, 
and Archeops connects the calibration of DMR to that of the many experiments sensitive to
the 1st acoustic peak and beyond.
The cluster normalization has dropped somewhat because of recalibrations 
of the mass-temperature relation (see Table 3).

Although the lensing normalization has been uniformly high across groups
\cite{Bacon02,Hoekstra02a,Refregier02,Waerbeke02}, the largest and most recent 
data set released has now produced the much lower normalization 
$\sigma_8\approx 0.71$ \cite{Jarvis02} (for $\Om=0.3$).
Indeed, the Hoekstra {\etal} normalization may
also be coming down by 5-10\% 
(Hoekstra, private communication), partly because of a new and more accurate version 
of the Peacock \& Dodds fitting formula \cite{PeacockDodds96}.

Moreover, our analysis did not marginalize over the source redshift 
parameters $\alpha=4.7$, $\beta=1.7$ and $z_0=0.302$ from \eq{SourceDistEq}, 
but merely used the best fit values from \cite{Hoekstra02a}.
Such marginalization would be expected to somewhat weaken the $\sigma_8$ constraints
that we obtained from the RCS lensing data.
As an extreme example, we recomputed the shear power spectrum $P_\kappa(\l)$ 
replacing the characteristic source redshift parameter $z_s=0.302$ with 
the $3-\sigma$ lower and upped limits on its value from \cite{Hoekstra02a}.
These extreme values
$z_s=z=0.274$ and $z_s=0.337$ made the shear fluctuation amplitude
$10-15\%$ lower and higher, respectively,
so lowering the source redshifts at the $2\sigma$ level would close approximately
half of the $\sigma_8$ gap between the RCS and the CMB.
Despite the current discord, there is thus real hope that 
a beautiful consistent picture of cosmology will eventually emerge.

\subsection{Are the constraints too good to be true?}

In addition to the above-mentioned tension regarding the best fit values of some parameters,
there is a slight puzzle regarding their error bars: some constraints seem a bit too good to be
true. Specifically, the CMB error bars in Table 2 are in some cases comparable to 
those forecast for MAP \cite{parameters2} using the Fisher matrix technique \cite{karhunen}.
This puzzling fact is not limited to the present paper, but appears rather generic.
For instance, consider the scalar spectral index $\ns$.
The forecast for MAP (2 years of data including polarization) is accuracy $\Delta\ns=0.11$
with no prior assumptions, $\Delta\ns=0.045$ assuming $\Ok=0$ and $\Delta\ns=0.021$ when 
adding SDSS galaxy clusterinng data (the final luminous red galaxy sample) \cite{parameters2}.
We found $\Delta\ns=0.06$ from current CMB alone (Table 2, assuming $\Ok=0$), comparable
to other studies in the recent literature.
For instance, Lewis \& Bridle \cite{Lewis02} obtain $\Delta\ns=0.04$ including 2dF data,
and the Boomerang team reports $\Delta\ns=0.06-0.08$ for various priors \cite{Ruhl02}.

\def\D{{\bf D}}
\def\F{{\bf F}}
\def\Fcmb{\F_{\rm cmb}}
\def\Flss{\F_{\rm 2df}}
\def\SS{{\bf\Sigma}}
\def\W{{\bf W}}

To gain insight into this puzzle, we compute the Fisher matrix \cite{karhunen} corresponding
to the {\it current} data. In the approximation that all the CMB information is 
coming from the mean rather than the covariance of the power, the equations in
\cite{karhunen} give
\beq{FisherEq}
\Fcmb = \D^t\W\SS^{-1}\W^t\D,
\eeq
where $\SS$ is the $28\times 28$ covariance matrix of our combined CMB power spectrum 
measurements from \sec{CMBdata}, $\W$ is the corresponding 
$\lmax\times 28$ window matrix and the $\lmax\times 28$ matrix 
\beq{DdefEq}
\D_{\l i}\equiv {\partial\over\partial p_i} \delta T_\l^2
\eeq
gives the derivatives of the power spectrum with respect to the cosmological parameters
$p_i$.
We compute an analogous Fisher matrix $\Flss$ using the covariance matrix and window functions
for the 2dF power spectrum from \cite{Tegmark2df01}. We then evaluate the attainable error bars 
from CMB alone as $\Delta p_i=(\Fcmb^{-1})_{ii}^{1/2}$ 
and 
from CMB+2dF as $\Delta p_i=([\Fcmb+\Flss]^{-1})_{ii}^{1/2}$.
For the scalar spectral index with an $\Ok=0$ prior, this gives
$\Delta\ns=0.16$ from CMB alone and $\Delta\ns=0.13$ for from CMB+2dF,
a factor 2-3 larger than the above-mentioned likelihood analysis results from us and others.

Comparing with our Fisher matrix results, the most suspicious error bar in Table 2
is that on the optical depth $\tau$, where we quoted $\Delta\tau=0.06$ from CMB alone
and the Fisher matrix gives $\Delta\tau=0.18$.
For the $\tau$ case, there is wide scatter in the literature.
On the optimistic side, the CBI team reported $\tau=0.09^{+.14}_{-0.07} $\cite{CBI02a},
and a strong detection of reionization was claimed from earlier data \cite{Schmalzing00}.
On the pessimistic side, the 2dF and Boomerang teams report a $2\sigma$ limit $\tau<0.5$ \cite{Efstathiou01,Ruhl02},
in line with the Fisher forecast.

If our $\tau$ errors are indeed too small due to some artifact of our method, 
this would weaken the constraints on the normalization $\sigma_8$, since CMB constrains mainly 
the combination $\sigma_8 e^{-\tau}$. However, the conclusion that $\sigma_8=0.9$ forces the uncomfortably 
high value $\tau\approx 0.2$ would still stand, since it follows directly from this scaling relation.


Efstathiou \& Bond \cite{EfstathiouBond98} have shown that there are physically understandable
reasons why the Fisher matrix approach can significantly overestimate the errors on cosmological parameters, 
especially when 
strong degeneracies are present and that the curvature of a banana-shaped allowed region
becomes important. This is indeed the case with $\tau$, with its strong degeneracy 
with the normalization parameter $\As$, and a rigorous Bayesean confidence interval taking 
into account the asymmetry $\tau\ge 0$ (neglected in the Fisher treatment) also matters.
``Weak'' priors can also have a strong effect, notably on $h$ and gravity waves.
Hopefully, our concerns will therefore go away when the MAP data arrives, in particular
when all major degeneracies are broken with other datasets and priors. 
Since \eq{FisherEq} is trivial to apply, it would be prudent to include a Fisher analysis of the 
data used in all future parameter constraint papers, as a reality check.

\subsection{Outlook}

In conclusion, the pace at which new measurements have tightened up constraints on cosmological parameters 
is breathtaking. The constraints become particularly tight if one uses theoretical prejudice to
impose spatial flatless $\Ok=0$ like we have done, interpreting the CMB indication $\Ok\approx 0$ as
support for inflation and perfect flatness and thereby breaking the worst parameter degeneracy of all 
by fiat.
It can be argued that precision cosmology is already here, before MAP, and now faces
its first baptism of fire: will the current precision measurements of cosmological parameters agree with MAP?
Next month we will know whether the whole edifice goes down in flames or stands stronger than ever.

The authors wish to thank Gary Bernstein, Karim Benabed and Roman Scoccimarro for helpful comments.
This work was supported by 
NSF grants AST-0071213, AST-0134999, AST-0098606 and PHY-0116590,
NASA grants NAG5-9194, NAG5-11099, NAG5-10923 and NAG5-10924 and a 
Keck foundation grant.
MT and MZ are David and Lucile Packard Foundation Fellows
MT is a Cottrell Scholar of Research Corporation.




\ed